\documentclass{acm_proc_article-sp}
\usepackage{url}
\usepackage[utf8x]{inputenc}

\begin{document}

\title{Visualizing criminal networks\\reconstructed from mobile phone records}

\numberofauthors{3}

\author{
Emilio Ferrara\\
\affaddr{School of Informatics and Computing}\\
\affaddr{Indiana University Bloomington, USA}\\ 
\email{\footnotesize ferrarae@indiana.edu}
\alignauthor
Pasquale De Meo\\
\affaddr{Department of Ancient and Modern Civilizations}\\
\affaddr{University of Messina, Italy}\\ 
\email{\footnotesize pdemeo@unime.it}
\\
\alignauthor
Salvatore Catanese, Giacomo Fiumara\\
\affaddr{Department of Mathematics and Computer Science}\\
\affaddr{University of Messina, Italy}\\ 
\email{\footnotesize \{scatanese,gfiumara\}@unime.it}
}

\maketitle
\begin{abstract}
In the fight against the racketeering and terrorism, knowledge about the structure and the organization of criminal networks is of fundamental importance for both the investigations and the development of efficient strategies to prevent and restrain crimes.
Intelligence agencies exploit information obtained from the analysis of large amounts of heterogeneous data deriving from various informative sources including the records of phone traffic, the social networks, surveillance data, interview data, experiential police data, and police intelligence files, to acquire knowledge about criminal networks and initiate accurate and destabilizing actions.
In this context, visual representation techniques coordinate the exploration of the structure of the network together with the metrics of  social network analysis.
Nevertheless, the utility of visualization tools may become limited when the dimension and the complexity of the system under analysis grow beyond certain terms.
In this paper we show how we employ some interactive visualization techniques to represent criminal and terrorist networks reconstructed from phone traffic data, namely foci, fisheye and geo-mapping network layouts.
These methods allow the exploration of the network through animated transitions among visualization models and local enlargement techniques in order to improve the comprehension of interesting areas.
By combining the features of the various visualization models it is possible to gain substantial enhancements with respect to classic visualization models, often unreadable in those cases of great complexity of the network.
\end{abstract}

\category{}{Information systems}{World Wide Web}[Social networks]
\category{}{Networks}{Network types}[Social media networks]


\section{Introduction}
Criminal Network Analysis allows to identify structure and flow of information among the members of a criminal network and to acquire the knowledge necessary to plan proactive and reactive interventions.
Among the more frequently used analytic techniques there is the mapping of interactions among the members of the organization and their activities by means of a graph \cite{mena2003investigative}.
A graph representation allows to overview the network structure, to identify the cliques, the groups, and the key players.
The possibility of mapping the attributes of data and metrics of the network using visual properties of the nodes and edges makes this technique a powerful investigative tool.
Often, however, visualization techniques become discouraging as a consequence of density and dimensions of the network.
Some obstacles such as the overlap of nodes and the dense intersections of edges severely reduce the readability of the graph.
In other words, there is a limit to the number of elements which can be distinctly viewed from the human eye.
An influential theory about the improvement of the quality of network visualization has been suggested by Shneiderman in \cite{Shneiderman2006}, where the so-called ``Network Nirvana'' is described.
According to this theory, some demanding targets must be pursued: i) the visibility of each node; ii) the possibility of counting the degree of each node; iii) the possibility of following each edge from the source to the destination nodes and, iv) the possibility of identifying the clusters.
Although it can be challenging, or even impossible, to satisfy all these conditions at the same time as the network grows in size and complexity, an effective network analysis strategy should try to optimize the visualization methods in order to incorporate these guidelines.
In this work, we present three visualization techniques that yield better network representations that, in turn, allow for enhanced data interpretability; we discuss these layout techniques, namely fisheye, foci and network geo-mapping, specifically in the context of criminal network analysis, but we do not exclude a broader applicability to other domains of social network analysis (SNA).

\subsection{Literature on criminal network analysis}
In the last thirty years academic research related to the application of social network analysis to intelligence and study of criminal organizations has constantly grown.
One of the most important studies is due to Malcolm Sparrow \cite{sparrow91}, related to the application of the techniques of analysis of networks, and their vulnerabilities, for intelligence scopes.

Sparrow defined four features peculiar of criminal networks (CNs), namely: i) limited dimension --- CNs are often composed of at most few thousand nodes; ii) information incompleteness --- criminal or terrorist networks are unavoidably incomplete due to fragmentary available information and erroneous information; iii) undefined borders --- it is difficult to determine all the relations of a node; and, iv) dynamics --- new connections imply a constant evolution of the structure of the network.

Thanks to Sparrow's work, other authors tried to study criminal networks using the tools of SNA.
For example, Baker and Faulkner \cite{BakerFaulkner93} studied illegal networks in the field of electric plants and Klerks \cite{Klerks01thenetwork} focused on criminal organizations in The Netherlands.
In 2001, Silke \cite{Silke01} and Brennan et al. \cite{Brannan01} acknowledged a slow growth in the fight against terrorism, and examined the state of the art in the field of criminal network analysis.

Arquilla and Ronfeldt \cite{Arquilla01} summarize prior research by introducing the concept of Netwar and its applicability to terrorism.
They illustrate the difference between social networks and CNs, demonstrating the great utility of network models to understand the nature of criminal organizations.

All these early studies somehow neglected the importance of network visualization, stressing aspects related more to statistical network characterization, or interpretation of individuals' roles rooted in social theory.
However, in 2006, a popular work by Valdis Krebs \cite{Krebs02mappingnetworks} applied graph analysis in conjunction with network visualization theory to analyze the Al Qaeda cell responsible of the 2001-09-11 terrorist attacks in the USA.
This work represents a starting point of a series of academic papers in which social network analysis methods become applied to a real-world cases, differently from previous work where mostly toy models and fictitious networks were used.
Krebs' paper is one of the more cited papers in the field of application of social network analysis to Criminal Networks and it inspired further research in network visualization for the design and development of better SNA tools applications to support intelligence agencies in the fight against terror, and law enforcement agencies in their quest fighting crime.

\section{The problem}
\label{sec:problem}

In criminology and research on terrorism, SNA has been proved a powerful tool to learn the structure of a criminal organization. It allows analysts to understand the structural relevance of single actors and the relations among members, when regarded as individuals or members of (one or more) subgroup(s).
SNA defines the key concepts to characterize network structure and roles, such as centrality \cite{freeman1977set}, node and edge betweenness \cite{freeman1977set,brandes2001faster,demeo2012novel}, and structural similarity \cite{Lorrain1971}.
The understanding of network structure derived from these concepts would not be possible otherwise \cite{wasserman1994social}. 
The above-mentioned structural properties are heavily employed to visually represent social and criminal networks as a support decision-making processes.

SNA provides key techniques including the possibility to detect clusters, identify the most important actors and their roles and unveil interactions through various graphical representation methodologies \cite{confYangLS06}. 
Some of these methods are explicitly designed to identify groups within the network, while others have been developed to show social positions of group members.
The most common graphical layouts have historically been the node-link and the matrix representations \cite{Freeman00}.

Visualization has become increasingly important to gain information about the structure and the dynamics of social networks: since the introduction of sociograms, it appeared clear that a deep understanding of a social network was not achievable only through some statistical network characterization \cite{wasserman1994social}.

For all these reasons, a number of different challenges in network visualization have been proposed \cite{Schneider2009}.
The study of network visualization focuses on the solution of the problems related to clarity and scalability of the methods of automatic representation.
The development of a visualization system exploits various technologies and faces some fundamental aspects such as: i) the choice of the layout; ii) the exploration dynamics; and, iii) the interactivity modes introduced to reduce the visual complexity.

Recent studies tried to improve the exploration of networks by adding views, user interface techniques and modes of interaction more advanced than the conventional node-link and force-directed \cite{fruchterman1991graph} layouts.
For example, in \emph{SocialAction} \cite{perer2006balancing} users are able to classify and filter the nodes of the network according to the values of their statistical properties.
In \textit{MatrixExplorer} \cite{Henry:2006} the node-link layout is integrated with the matrix layout.
Nonetheless, these visualization systems have not been explicitly developed with the aim of the exhaustive comprehension of all properties of the network.
Users need to synthesize the results coming from some views and assemble metrics with the overall structure of the network.

Therefore, we believe that an efficient method to enhance the comprehension and the study of social networks, and in particular of criminal networks, is to provide a more explicit and effective node-link layout algorithm.
This way, important insights could be obtained from a unique layout rather than from the synthesis derived from some different layouts.

We recently presented a framework, called \emph{LogAnalysis} \cite{catanese2013forensic,ferrara2014detecting}, that incorporates various features of social network analysis tools, but explicitly designed to handle criminal networks reconstructed from phone call interactions. 
This framework allows to visualize and analyze the phone traffic of a criminal network by integrating the node-link layout representation together with the navigation techniques of zooming and focusing and contextualizing. The reduction of the visual complexity is obtained by using hierarchical clustering algorithms.
In this paper we discuss three new network layout methods that have been recently introduced in \emph{LogAnalysis}, namely fisheye, foci and geo-mapping, and we explain how these methods help investigators and law enforcement agents in their quest to fight crime.

It's worth noting that various tools to support network analysis exist. However, only few of them have been developed specifically for criminal network investigations.
We mention, among others, commercial tools like COPLINK \cite{chen2003coplink,xu2005crimenet}, Analyst's Notebook\footnote{\url{ibm.com/software/products/analysts-notebook/}}, Xanalysis Link Explorer\footnote{\url{http://www.xanalys.com/products/link-explorer/}} and Palantir Government\footnote{\url{http://www.palantir.com/solutions/}}. 
Other prototypes described in academic papers include Sandbox \cite{Brian2006} and POLESTAR \cite{Everett2006}. Some of these tools show similar features to \emph{LogAnalysis}, but, to the best of our knowledge, none of them yields the same effective and scalable network visualization with support to criminal networks reconstructed from phone call records.

\subsection{Aspects of structural analysis}
A central node of a criminal network may play a key role by acting as a leader, issuing orders, providing regulations or by effectively assuring the flow of information through the various components of the CN.
The removal of these central nodes may efficiently fragment the organization and interrupt the prosecution of a criminal activity.

Apart from studying the roles of various members, investigative officers must pay particular attention to subgroups or gangs each of which may be in charge of specific tasks.
Members of the organization must interact and cooperate in order to accomplish their illicit activities.
Therefore, the detection of subgroups whose members are tightly interrelated may increase the comprehension of the organization of the CN.
Moreover, groups may interact according to certain schemes.
For example, the members of a clan could frequently interact with the members of another and seldom with the remaining members of the network.
The detection of interaction models and the relations among the subgroups highlights information particularly useful about the overall structure of the network.

A significant aspect of the analysis of criminal networks is that it requires, differently from other networks, the ability of integrating information deriving from other sources in order to precisely understand its structure, operation and flow of information.
A typical process employed by an investigator is to start from one, or a few, known entities; after analyzing the associations these entities have with others, if any interesting association emerges, one may follow such a lead and keep expanding the associations until any significant link is uncovered between seemingly unrelated entities.

Mobile phone networks and online platforms are constantly used to perform or coordinate criminal activities \cite{xu2005criminal, morselli2010assessing}. 
Phone networks can be used to connect individuals involved in criminal activities in real time, often during real-world criminal events, from simple robberies to terror attacks.
Online platforms, instead, can be exploited to carry out illicit activities such as frauds, identity thefts or to access classified information.

The analysis of a criminal network is thus aimed at uncover the structural schemes of the organization, its operations and, even more importantly, the flow of communications among its members.
In modern investigative techniques the analysis of phone records represents a first approach that precedes a more refined scrutiny covering financial transactions and interpersonal relations.
For these reasons a structured approach is needed.

\begin{figure}[!t]
\centering
\includegraphics[width=\columnwidth]{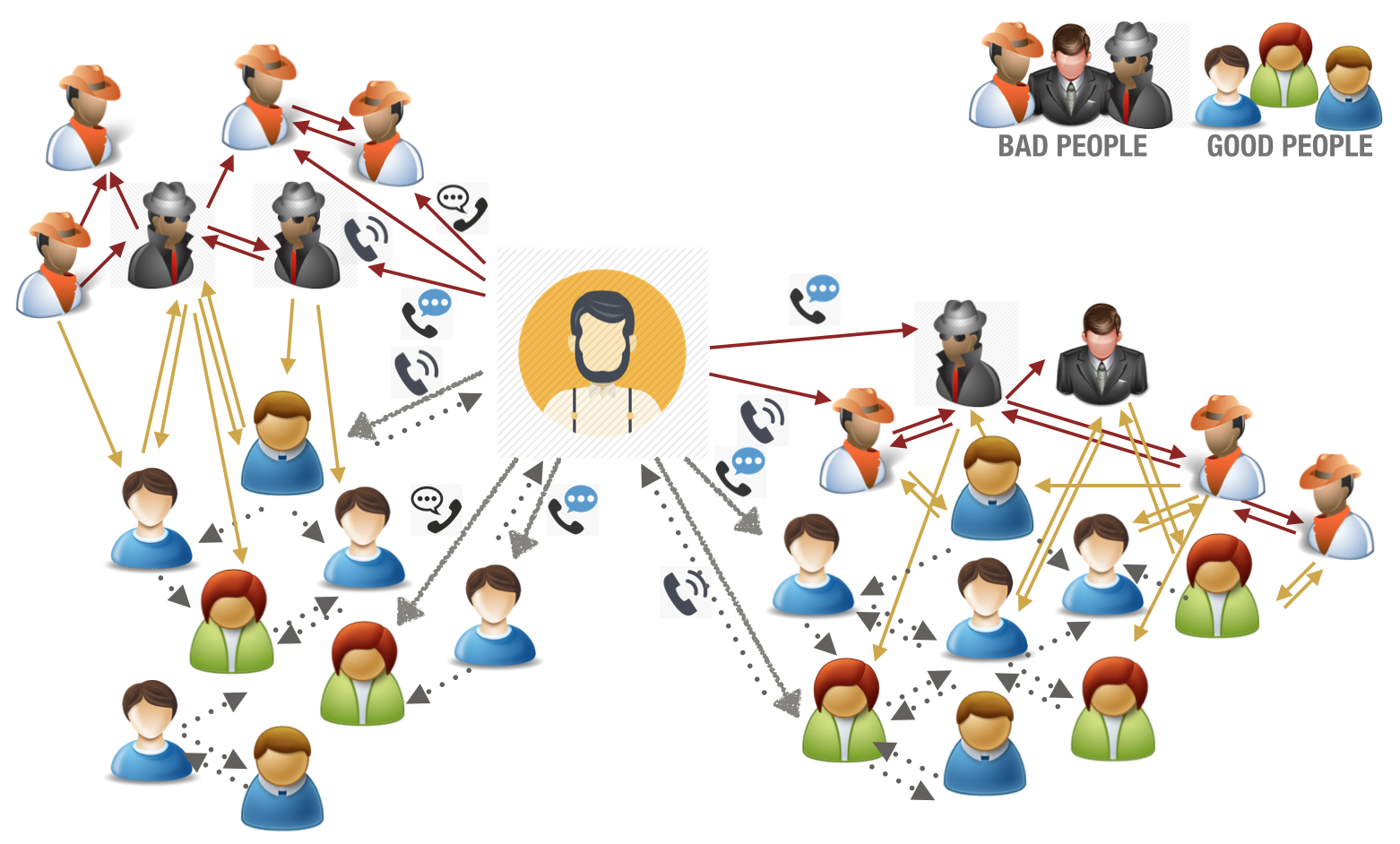}
\caption{Phone calls network of a suspected. Investigators start from some known entities, analyze the associations they have with others and expanding the associations until some significant link is uncovered. Here are highlighted personal interactions (gray arrows), links between criminal and personal connections of the suspect (yellow) and connections between members of the organization (in red).}
\label{fig:schema}
\end{figure}

Figure \ref{fig:schema} shows a stylized representation of a criminal network reconstructed from phone call records. We show the flow of phone communications of an individual subject of investigation, and we highlight various kind of phone interactions among individuals belonging to that person's social circles, and those belonging to the same criminal organization the individual is part of.

In the following we discuss three techniques that allow to efficiently and scalably inspect criminal networks reconstructed from phone interactions.

\begin{figure*}[!t]
\includegraphics[width=2.1\columnwidth]{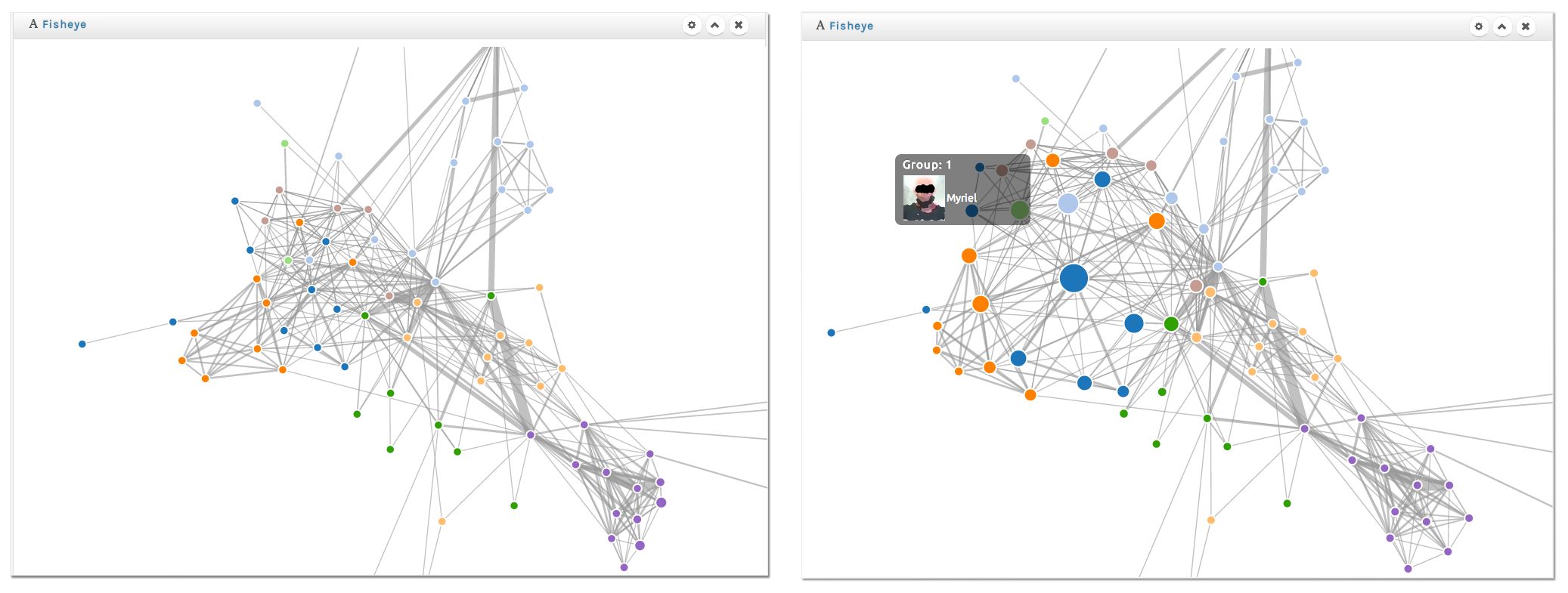}
\caption{The left picture shows a force-directed layout of a criminal network. On the right we depict the fisheye view of the same graph using transformation with distortion.}
\label{fig:fisheye}
\end{figure*}

\section{Visualization techniques}
\label{sec:viztech}

Typical network visualization tools rely on the popular force-directed layout \cite{fruchterman1991graph}.
The force-directed model represents the structure of the graph on the same foot as a physical system, in which nodes are physical points subject to various forces; nodes' coordinates (and therefore the layout itself) derive from the search of an equilibrium configuration of the physical system modeled by the algorithm \cite{Brandes2001}.
This particular layout arrangement has the advantage of grouping users in clusters which can be identified according to the heightened connectivity.
The Barnes-Hut algorithm \cite{barnes:hierarchical} associated to this layout simulates a repulsive N-body system in order to continuously update the position of the elements.

To optimize the visualization, it is possible to interactively modify the parameters relative to the tension of the springs (edges).
Nodes with low degree are associated a small tension and the elements are located in peripheral positions with respect to high degree nodes.
Other parameters can be tuned, such as spring tension, gravitational force and viscosity.
Our goal, in the following, is to suggest two methods to improve force-directed based layouts.
As we will show, these techniques are especially well suited for criminal network analysis; however, they could potentially be generalized for broader usage in other domains of network analysis --- for example, for applications in social and political sciences.

\subsection{Focus and context based visualization}

The number of edges within a network usually grows faster than the number of nodes.
As a consequence, the network layout would necessarily contain groups of nodes in which some local details would easily become unreadable because of density and overlap of the edges. As the size and complexity of the network grow, eventually nodes and edges become indistinguishable. 
This problem is known as visual overload \cite{Assa1997}. 
A commonly used technique to work around visual overload consists of employing a zoom-in function able to enlarge the part of the graph of interest. The drawback of this operation is the detriment of the visualization of the global structure which, during the zooming, would not be displayed.
However, such a compromise is reasonable in a number of situations including, in some cases, the domain of criminal network analysis. 

During an investigation, it is crucial to narrow down the analysis to the relevant suspects, to efficiently employ human and computational resources. Police officers typically draw some hypotheses about an individual suspect of being part of a criminal organization, or of being involved (or about to) in some crime; they concentrate the initial investigation on this individual, and on that person's social circles, as a ground to build the social network object of analysis.
The main role of visual analysis lies in allowing the detection of unknown relations, on the base of the available limited information.
A typical procedure starts from known entities, to analyze the relations with other subjects and continue to expand the network inspecting first the edges appearing the most between individuals apparently unrelated.
During this procedure, only some nodes are relevant and it is important to focus on them rather than on the network as a whole.

Nevertheless, a spring embedded layout (including force-directed ones) does not provide any support to this kind of focus and analysis.
In these situations, \emph{focus and context} visualization techniques are needed in order to help a user to explore a specific part of a complex network. To this purpose, we here introduce the fisheye and the foci layouts.

\subsection{Fisheye layout}

Focus and context is an interactive visualization technique \cite{Leung:1994}. 
It allows the user to focus on one or more areas of a social network, to dynamically tune the layout as a function of the focus, and to improve the visualization of the neighboring context.
The \emph{fisheye view} is a particular focus and context visualization technique which has been applied to visualize self-organizing maps in the Web surfing \cite{Yang:2003}.
It was first proposed by Furnas \cite{Furnas:1986} and successively enriched by Brown et al. \cite{SarkarB94}. It is known as a visualization technique that introduces distortion in the displayed information.

The fisheye layout is a local linear enlargement technique that, without modifying the size of the visualization canvas, allows to enhance the region surrounding the focus, while compressing the remote neighboring regions. The overall structure of the network is nevertheless maintained. An example of application of this technique is show in Figure \ref{fig:fisheye}. The picture shows a moderately small criminal network reconstructed from phone call interactions of about 75 individuals. The layout on the left panel is obtained by using a force-directed method implemented in our framework, \emph{LogAnalysis}. The analyst can inspect the nodes of the network, which contains known criminals, suspects, and their social circles. When the focus is applied on a given node, the visualization transitions to the fisheye layout (see the right panel). A tool-tip with additional information about the node appears when the node is selected --- it shows the phone number, personal details, address, photo, etc. The layout causes edges among remote nodes to experience stronger distortions than local nodes. The upside of the presented method is the possibility to achieve the three recommendations of Network Nirvana \cite{Schneider2009} when focusing on a given node: all the nodes' neighbors are clearly visible, the node degree is easily countable, and the edges incident on that node can be identified and followed.

Note that fisheye and force-directed layouts can be used in a complementary way.
By combining the two methods, our framework efficiently yields focus and context views.

\subsection{Foci layout}

\begin{figure}[!t]
\centering
\includegraphics[width=\columnwidth]{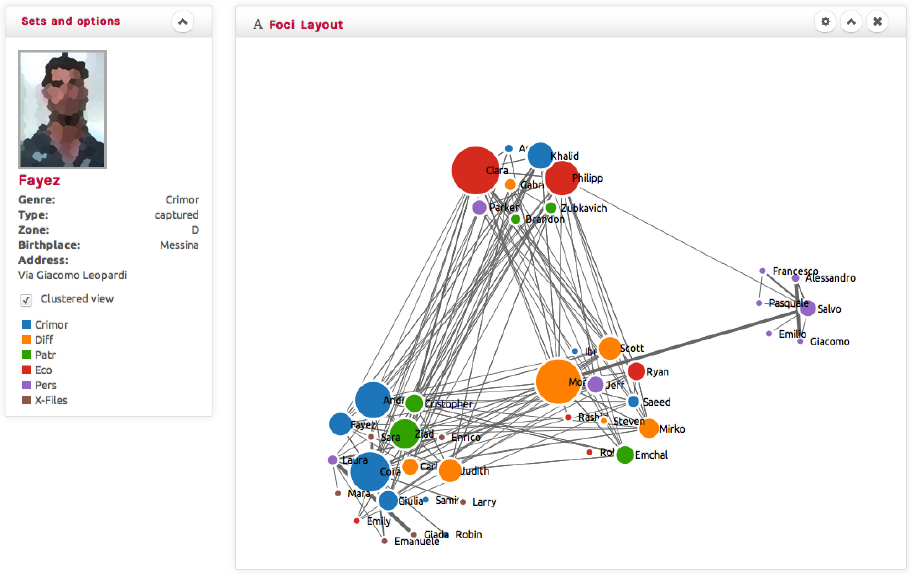}
\caption{Foci layout.}
\label{fig:foci}
\end{figure}

The \emph{foci layout} implements three network visualization models: force-directed, semantic and clustered layouts. The latter is based on the Louvain community detection algorithm \cite{Blondel2008,demeo2011generalized}. Future implementations will explore other methods \cite{demeo2013enhancing,demeo2014mixing}.
Our model supports multilayer analysis of the network through interactive transitions from the force-directed layout, with a single gravitational center, to the clustered one with more force centers placed in predetermined distinct areas.
This layout allows to analyze the network on various layering levels depending on specified node attributes.
Figure \ref{fig:foci} shows the phone traffic network of some clans the previous criminal network, in which the color of the nodes denotes the type of crime committed by the members.

In this example, the clustering truthfully reflects the known territorial division among the groups belonging to the organization. In Figure \ref{fig:foci} the focus is on a specific node.
Using this layout it is possible to contextually analyze the community structure, the type of committed crime in respect to the members of the clan, and the direct relations of each single individual. This layout integrates also the forth Network Nirvana recommendation, namely the possibility to identify clusters and to highlight the community structure.

\subsection{Network geo-mapping}

It is possible to extend the phone traffic analysis to include the phone logs recorded by the BTS (Base Transceiver Station), in which the GPS coordinates of the cell are reported.
All base stations are provided with directional antennas and each cell has two or more sectors. 
For each cell it is known the azimuth (direction) corresponding to the central axis of each sector, together with the width of the beam of each antenna, which determines the coverage angle of the sector.
These data do not allow to localize the geo-referenced position of the phones involved in the events recorded in the logs.
Nevertheless, it is possible, within a certain approximation, to localize the users falling within the coverage area.

Zang et al. \cite{Zang2010} described a technique based on Bayesian interference to localize mobile phones using additional information, such as the round-trip-time of data transmission packets and the measure of SINR (Signal to Interference plus Noise Ratio). 
The parameters obtained experimentally have been compared with the records of phone calls and the corresponding GPS entries to ascertain their distribution.
This localization technique produces satisfactory results with a reduction of the error amounting to a $20\%$ with respect to the \emph{blind approach}.
Traag et al. in \cite{Traag2011} used Bayesian interference to deduce, starting from phone traffic data, profiles about the places and the proximity of a given social event.

\begin{figure}[!t]
\centering
\includegraphics[width=\columnwidth]{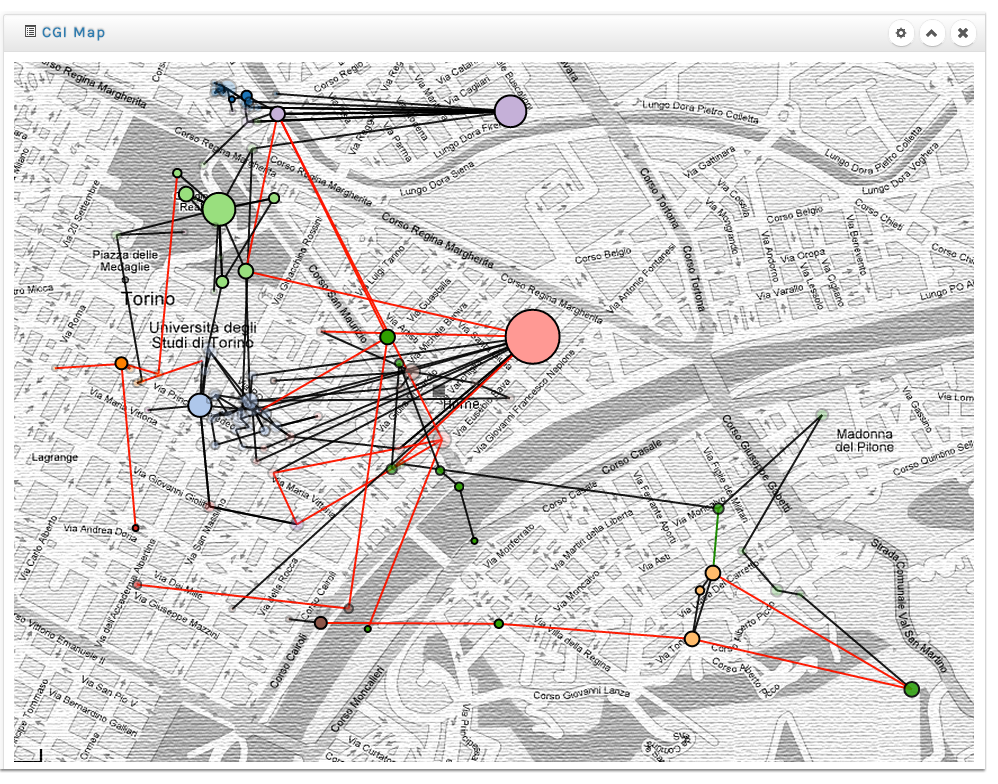}
\caption{Geo-mapping layout.}
\label{fig:geo}
\end{figure}

Our framework provides network geo-mapping by using this type of techniques to infer the spatial origin of each call. We here describe the network geo-mapping visualization method adopted in \emph{LogAnalysis}.
This layout allows to simultaneously carry out spatial and temporal relational analysis of phone call logs.
It places nodes of the network on a map, in correspondence of the coordinates of the cells linked during the events recorded in the logs.
Nodes are connected by links related to displacements.
Contacts falling within the sectors of a given zone are represented with nodes of the same color.
Information about displacements, routines and areas of interest for the investigation are displayed. The adoption of network geo-mapping has proved extremely useful during real investigations.
Figure \ref{fig:geo} shows, as an example, a case study in which larger nodes identify zones in which, in the time period of the investigation, a high number of contacts has been recorded among some members of the CN.
Unsurprisingly, the inspection by police officers of such high-profile locations provided crucial insights on the investigation. Unfolding the temporal evolution of the geo-mapped phone traffic network also allows to reproduce individuals' movements and communication dynamics during specific criminal events embedded in space and time, like robberies, assaults, or homicides.

\section{Conclusions}
\label{sec:conclusions}

Criminal network analysis benefits from visualization methods used to support the investigations, especially when dealing with networks reconstructed from heterogeneous data sources, characterized by increasing size and complexity.
In this paper we integrated the spring embedded algorithm with the fisheye and foci layouts to allow interactive exploration of criminal networks through our network analysis framework.
The combination of these techniques proved helpful to support investigators in the extraction of useful information and critical insights, to identify key members in terrorist groups, and to discover specific paths of interaction among members of criminal organizations.
Experimental results show that the combination of force-directed layouts, distortion techniques and multi-force systems yield better performance in terms of both efficiency and efficacy.

\footnotesize
\bibliographystyle{abbrv}
\bibliography{sigproc}  

\begin{thebibliography}{10}

\bibitem{Arquilla01}
J.~Arquilla and D.~Ronfeldt.
\newblock Networks and netwars: The future of terror, crime, and militancy.
\newblock {\em Survival}, 44(2):175--176, 2001.

\bibitem{Assa1997}
J.~Assa, D.~Cohen-Or, and T.~Milo.
\newblock Displaying data in multidimensional relevance space with 2d
  visualization maps.
\newblock In {\em Proc. Visualization '97}, pages 127--134, 1997.

\bibitem{BakerFaulkner93}
W.~Baker and R.~Faulkner.
\newblock The social organization of conspiracy: illegal networks in the heavy
  electrical equipment industry.
\newblock {\em Am. Social. Rev.}, 58, 1993.

\bibitem{barnes:hierarchical}
J.~Barnes and P.~Hut.
\newblock A hierarchical o(n log n) force calculation algorithm.
\newblock {\em Nature}, 324:446--449, 1986.

\bibitem{Blondel2008}
V.~D. Blondel, J.-L. Guillaume, R.~Lambiotte, and E.~Lefebvre.
\newblock Fast unfolding of communities in large networks.
\newblock {\em Journal of Statistical Mechanics: Theory and Experiment},
  2008(10):P10008+, July 2008.

\bibitem{brandes2001faster}
U.~Brandes.
\newblock {A faster algorithm for betweenness centrality}.
\newblock {\em Journal of Mathematical Sociology}, 25(2):163--177, 2001.

\bibitem{Brandes2001}
U.~Brandes.
\newblock Drawing on physical analogies.
\newblock In {\em Drawing Graphs}, pages 71--86. Springer, 2001.

\bibitem{Brannan01}
D.~W. Brannan, P.~F. Esler, and N.~T. Anders~Strindberg.
\newblock Talking to terrorists: Towards an independent analytical framework
  for the study of violent substate activism.
\newblock {\em Studies in Conflict and Terrorism}, 24(1):3--24, 2001.

\bibitem{catanese2013forensic}
S.~Catanese, E.~Ferrara, and G.~Fiumara.
\newblock Forensic analysis of phone call networks.
\newblock {\em Social Network Analysis and Mining}, 3(1):15--33, 2013.

\bibitem{chen2003coplink}
H.~Chen, D.~Zeng, H.~Atabakhsh, W.~Wyzga, and J.~Schroeder.
\newblock Coplink: managing law enforcement data and knowledge.
\newblock {\em Comm. ACM}, 46(1):28--34, 2003.

\bibitem{demeo2011generalized}
P.~De~Meo, E.~Ferrara, G.~Fiumara, and A.~Provetti.
\newblock {Generalized Louvain method for community detection in large
  networks}.
\newblock In {\em Proc. 11th International Conference on Intelligent Systems
  Design and Applications}, pages 88--93. IEEE, 2011.

\bibitem{demeo2013enhancing}
P.~De~Meo, E.~Ferrara, G.~Fiumara, and A.~Provetti.
\newblock Enhancing community detection using a network weighting strategy.
\newblock {\em Information Sciences}, 222:648--668, 2013.

\bibitem{demeo2014mixing}
P.~De~Meo, E.~Ferrara, G.~Fiumara, and A.~Provetti.
\newblock Mixing local and global information for community detection in large
  networks.
\newblock {\em Journal of Computer and System Sciences}, 80(1):72--87, 2014.

\bibitem{demeo2012novel}
P.~De~Meo, E.~Ferrara, G.~Fiumara, and A.~Ricciardello.
\newblock A novel measure of edge centrality in social networks.
\newblock {\em Knowl-based Syst}, 30:136--150, 2012.

\bibitem{ferrara2014detecting}
E.~Ferrara, P.~De~Meo, S.~Catanese, and G.~Fiumara.
\newblock Detecting criminal organizations in mobile phone networks.
\newblock {\em Expert Systems with Applications}, 41(13):5733--5750, 2014.

\bibitem{freeman1977set}
L.~Freeman.
\newblock A set of measures of centrality based on betweenness.
\newblock {\em Sociometry}, pages 35--41, 1977.

\bibitem{Freeman00}
L.~C. Freeman.
\newblock Visualizing social networks.
\newblock {\em Journal of Social Structure}, 1, 2000.

\bibitem{fruchterman1991graph}
T.~Fruchterman and E.~Reingold.
\newblock {Graph drawing by force-directed placement}.
\newblock {\em Software: Practice and Experience}, 21(11):1129--1164, 1991.

\bibitem{Furnas:1986}
G.~W. Furnas.
\newblock Generalized fisheye views.
\newblock {\em SIGCHI Bull.}, 17(4):16--23, Apr. 1986.

\bibitem{Henry:2006}
N.~Henry and J.-D. Fekete.
\newblock Matrixexplorer: A dual-representation system to explore social
  networks.
\newblock {\em IEEE Transactions on Visualization and Computer Graphics},
  12(5):677--684, Sept. 2006.

\bibitem{Klerks01thenetwork}
P.~Klerks and E.~Smeets.
\newblock The network paradigm applied to criminal organizations: Theoretical
  nitpicking or a relevant doctrine for investigators?
\newblock {\em Connections}, 24:53--65, 2001.

\bibitem{Krebs02mappingnetworks}
V.~Krebs.
\newblock Mapping networks of terrorist cells.
\newblock {\em Connections}, 24(3):43--52, 2002.

\bibitem{Leung:1994}
Y.~K. Leung and M.~D. Apperley.
\newblock A review and taxonomy of distortion-oriented presentation techniques.
\newblock {\em ACM Trans. Comput.-Hum. Interact.}, 1(2):126--160, June 1994.

\bibitem{Lorrain1971}
F.~Lorrain and H.~C. White.
\newblock Structural equivalence of individuals in social networks.
\newblock {\em The Journal of Mathematical Sociology}, 1(1):49--80, 1971.

\bibitem{mena2003investigative}
J.~Mena.
\newblock {\em Investigative Data Mining for Security and Criminal Detection}.
\newblock Butterworth-Heinemann, 2003.

\bibitem{morselli2010assessing}
C.~Morselli.
\newblock Assessing vulnerable and strategic positions in a criminal network.
\newblock {\em Journal of Contemporary Criminal Justice}, 26(4):382--392, 2010.

\bibitem{perer2006balancing}
A.~Perer and B.~Shneiderman.
\newblock {Balancing systematic and flexible exploration of social networks}.
\newblock {\em IEEE Trans. Visual. and Computer Graphics}, pages 693--700,
  2006.

\bibitem{Everett2006}
N.~J. Pioch and J.~O. Everett.
\newblock Polestar: collaborative knowledge management and sensemaking tools
  for intelligence analysts.
\newblock In {\em Proc. 15th ACM international conference on Information and
  knowledge management}, pages 513--521. ACM, 2006.

\bibitem{SarkarB94}
M.~Sarkar and M.~H. Brown.
\newblock Graphical fisheye views.
\newblock {\em Comm. ACM}, 37(12):73--84, 1994.

\bibitem{Schneider2009}
F.~Schneider, A.~Feldmann, B.~Krishnamurthy, and W.~Willinger.
\newblock {Understanding online social network usage from a network
  perspective}.
\newblock In {\em Proc. 9th SIGCOMM conference on Internet measurement
  conference}, pages 35--48. ACM, 2009.

\bibitem{Shneiderman2006}
B.~Shneiderman and A.~Aris.
\newblock Network visualization by semantic substrates.
\newblock {\em IEEE Trans. Visual. and Computer Graphics}, 12(5):733--740, Sept
  2006.

\bibitem{Silke01}
A.~Slike.
\newblock The devil you know: Continuing problems with research on terrorism.
\newblock {\em Terrorism and Political Violence}, 13:1--14, 2001.

\bibitem{sparrow91}
M.~K. Sparrow.
\newblock The application of network analysis to criminal intelligence: An
  assessment of the prospects.
\newblock {\em Social Networks}, 13(3):251--274, 1991.

\bibitem{Traag2011}
V.~A. Traag, A.~Browet, F.~Calabrese, and F.~Morlot.
\newblock Social event detection in massive mobile phone data using
  probabilistic location inference.
\newblock In {\em 2011 IEEE 3rd international conference on social computing
  (socialcom)}, pages 625--628. IEEE, 2011.

\bibitem{wasserman1994social}
S.~Wasserman and K.~Faust.
\newblock {\em Social network analysis: methods and applications}.
\newblock Cambridge Univ. Pr., 1994.

\bibitem{Brian2006}
W.~Wright, D.~Schroh, P.~Proulx, A.~Skaburskis, and B.~Cort.
\newblock The sandbox for analysis: concepts and methods.
\newblock In {\em Proc. SIGCHI Conference on Human Factors in Computing
  Systems}, pages 801--810, 2006.

\bibitem{xu2005crimenet}
J.~Xu and H.~Chen.
\newblock Crimenet explorer: a framework for criminal network knowledge
  discovery.
\newblock {\em ACM Trans. on Information Systems}, 23(2):201--226, 2005.

\bibitem{xu2005criminal}
J.~Xu and H.~Chen.
\newblock Criminal network analysis and visualization.
\newblock {\em Comm. ACM}, 48(6):100--107, 2005.

\bibitem{Yang:2003}
C.~Yang, H.~Chen, and K.~Hong.
\newblock Visualization of large category map for internet browsing.
\newblock {\em Decis. Support Syst.}, 35(1):89--102, Apr. 2003.

\bibitem{confYangLS06}
C.~Yang, N.~Liu, and M.~Sageman.
\newblock Analyzing the terrorist social networks with visualization tools.
\newblock In {\em Intelligence \& security informatics}. 2006.

\bibitem{Zang2010}
H.~Zang, F.~Baccelli, and J.~Bolot.
\newblock Bayesian inference for localization in cellular networks.
\newblock In {\em 2010 Proceedings IEEE INFOCOM}, pages 1--9. IEEE, 2010.

\end{thebibliography}
\balancecolumns
\end{document}